\documentclass[12pt,preprint]{aastex}

\usepackage{graphicx}

\title{A VLBI polarization study of SiO masers towards VY CMa}

\author{L.L. Richter}
\affil{Physics and Electronics Department, Rhodes University, Grahamstown, South Africa}

\author{A.J. Kemball}
\affil{National Center for Supercomputing Applications, University of Illinois at
Urbana-Champaign, USA}

\author{J.L. Jonas}
\affil{Physics and Electronics Department, Rhodes University, Grahamstown, South Africa}

\begin{document}

\begin{abstract}
Maser emission from the SiO molecule has been widely observed in the
near-circumstellar envelopes of late-type, evolved stars. VLBI images can
resolve individual SiO maser spots, providing information about the
kinematics and magnetic field in the extended atmospheres of these stars.
This poster presents full polarization images of several SiO maser lines
towards the supergiant star VY CMa. VY CMa is a particularly strong SiO
maser source and allows observations of a wide range of maser transitions.
We discuss implications of these observations for VY CMa morphology,
polarization, and pumping models.
\end{abstract}

\section{Introduction}

SiO maser emission has been widely observed in the near-circumstellar envelopes of late-type 
evolved stars. VLBI images can resolve individual SiO maser spots around these stars, 
providing information about the kinematics and magnetic field close to the stellar surface 
at high spatial resolution.
Comparison of the emission from different SiO maser lines may provide insight into the 
pumping mechanisms providing the population inversion necessary for the masers and 
also holds promise in discriminating between competing SiO maser polarization models.

\section{Observations}

The source VY CMa was observed with the Very Long Baseline Array (VLBA) on 2 December 1994 
(epoch 1) and 20 and 23 December 2003 (epoch 2). The data were reduced and imaged with AIPS 
following the technique described by \citet{Kemball:95}.
Several SiO maser lines were observed, three of which are shown in Figure \ref{fig:maps}.
Unfortunately there was very little data for each of the epoch 2 lines 
and the signal to noise ratio is lower than it is in epoch 1.

\section{Discussion}

\begin{figure}
\includegraphics[width=5.3in,angle=0]{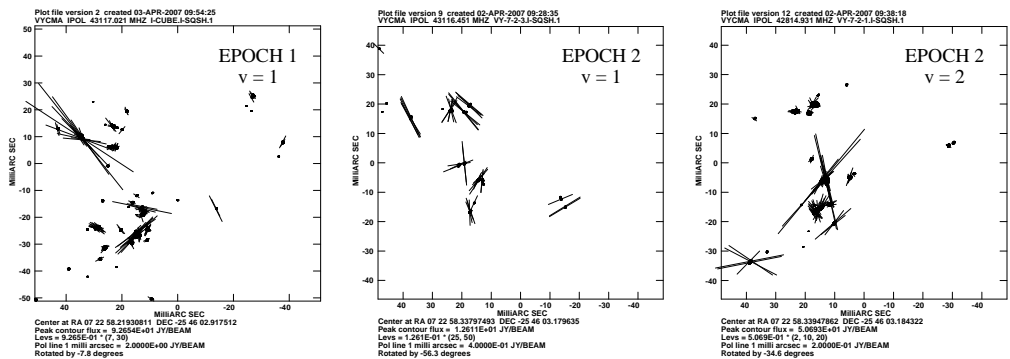}
\caption{43 GHz J=1-0 SiO masers towards VY CMa. The intensity is given by the contours and 
the magnitude and direction of the linear polarisation are represented the vectors.}
\label{fig:maps}
\end{figure}

The intensity of the imaged maser emission in epoch 1 was significantly higher than the 
intensity of any of the epoch 2 lines.
The epoch 2 v=1 J=1-0 and v=2 J=1-0 emission has a very similar spatial distribution. 
These maps in Figure \ref{fig:maps} were aligned by correlating the images, channel by 
channel.
There is no obvious trend in the direction of the polarization vectors between epochs.

\subsection{Pumping models}

The primary pumping mechanism driving SiO maser emission remains an issue of dispute in the 
literature. Models of masers in the circumstellar environment make use of both radiative 
pumping \citep[e.g.][]{Bujar:94} and collisional pumping \citep[e.g.][]{Lockett:92, Doel:95}. 

One means to identify the primary pumping mechanism at work is to compare the spatial distribution 
of different vibrationally-excited SiO maser lines. The spatial coincidence of the rotational 
transitions in different vibrational lines, such as the v=1 and v=2 J=1-0 lines, would argue 
against a radiative pumping model \citep{Bujar:94, Doel:95}. 
Kinematic models of the SiO maser emission, using collisional pumping, predict that the v=1 J=1-0 
emission should lie further from the star than the v=2 J=1-0 emission, in a thicker shell
\citep{Gray:00}.

In the epoch 2 maps above, the v=1 J=1-0 and v=2 J=1-0 maps have many overlapping features and a 
similar overall distribution. The overlap of several features in the aligned v=2 J=1-0 and v=1 
J=1-0 maps argues against purely radiative pumping. Where the v=1 and v=2 features overlap, the 
v=2 features generally extend inwards further than the v=1 features. 

We cannot draw definitive conclusions from comparisons of different rotational lines at just one 
epoch. Further observations in this area are continuing.

\begin{acknowledgments}
I gratefully acknowledge the support of the NRF, the Fuchs Foundation and the Rhodes University 
Physics Department.
\end{acknowledgments}


\begin{thebibliography}{}

\bibitem[Bujarrabal, 1994]{Bujar:94}
     Bujarrabal, V. 1994,
     \textit{A\&A} 285, 953   
    
\bibitem[Doel et al., 1995]{Doel:95} 
     Doel, R.C., Gray, M.D., Humphreys, E.M.L., Braithwaite, M.F. \& Field, D. 1995, 
     \textit{A\&A} 302, 797
     
\bibitem[Gray \& Humphreys, 2000]{Gray:00}
     Gray, M.D. \& Humphreys, E.M.L. 2000,
     \textit{New Astron.} 5, 155
     
\bibitem[Kemball \& Diamond, 1995]{Kemball:95}
    Kemball, A.J., Diamond, P.J. \& Cotton, W.D. 1995,
    \textit{A\&AS} 110, 383
    
\bibitem[Lockett \& Elitzur, 1992]{Lockett:92}
     Lockett, P. \& Elitzur, M. 1992, 
     \textit{ApJ}, 399, 704

\end{thebibliography}
\end{document}